\documentclass[12pt,notoc,nohyper]{ACLS}
\usepackage{amssymb}
\usepackage{epsfig,multicol,bbm}

\makeatletter
\def\blfootnote{\xdef\@thefnmark{Hi my dear}\@footnotetext}
\makeatother

\newcommand\fverb{\setbox\pippobox=\hbox\bgroup\verb}
\newcommand\fverbdo{\egroup\medskip\noindent            \fbox{\unhbox\pippobox}\ }
\newcommand\fverbit{\egroup\item[\fbox{\unhbox\pippobox}]}
\newbox\pippobox
\preprint{}

\title{Accelerating black holes, spin-3/2 fields and C-metric}
\author{Hai Lin$^{2,3}$, K. Saifullah$^{1,2}$, Shing-Tung Yau$^{2,3}$ \\
$^1$Department of Mathematics, Quaid-i-Azam University, Islamabad,
Pakistan
\\
$^2$Department of Physics, and Center for the Fundamental Laws of Nature,\\
~Harvard University, MA 02138, USA \\
$^3$Department of Mathematics, Harvard University, MA 02138, USA \\
}

\abstract{We consider spin-3/2 particles on the background of
general accelerating black holes and $C$-metric. The
Rarita-Schwinger equations of spin-3/2 particles are analyzed on
these backgrounds. The emission and absorption probabilities of
the spin-3/2 particles on these spacetimes are calculated. These
backgrounds which we analyze contain both black hole horizon and
acceleration horizon, and have general electric and magnetic
charges, rotation, and acceleration parameter. The properties of
the spin-3/2 field near the acceleration horizon are also
computed.
 }
\input{tcilatex}

\begin{document}


\section{Introduction}

\label{sec_introduction}

In this paper, we consider spin-3/2 particles on the spacetime of
accelerating black holes. The accelerating black hole is a black hole with
the source that has an acceleration \cite{Griffiths:2006tk, Griffiths:2005mi}%
. These black holes have an acceleration parameter and can also have charge
and rotation. In the coordinate system convenient for a boost-rotation
symmetry, the extended space can describe a pair of black holes accelerating
in opposite directions \cite{Griffiths:2006tk, Griffiths:2005mi, Bicak,
Pravda:2000zm, Corn Utt}. This space is the $C$-metric, and it can also
include charge and rotation \cite{Griffiths:2006tk, Griffiths:2005mi}.

The spin-3/2 particles can be described by the Rarita-Schwinger equations
\cite{Rarita Schwinger}. The spin-3/2 field can appear as effective fields
of spin-3/2 baryons, for example hadronic resonances, in effective field
theories of interacting hadrons \cite{Benmerrouche:1989uc, Hemmert:1997ye}.
It can also appear as gravitinos in supergravities in which there is
fermionic gauge invariance, for example \cite{Rindani:1986tj, Ferrara:1992yc}%
, and their dimensional reductions, for example \cite{Rindani:1986tj}. The
spin-3/2 particle is also of interest for phenomenological models beyond the
standard model.

One the other hand, quantum mechanical effects of particles on the
background of general relativity can give rise to many interesting phenomena
such as the evaporation of black holes via Hawking radiations \cite%
{Hawking:1974sw, Gibbons:1977mu}. These radiations have also been
investigated as quantum tunneling of particles from the black hole
horizons, for example \cite{Kraus:1994fh, Parikh:1999mf}. Many
black holes, such as the Reissner-Nordstr\"{o}m and Kerr-Newman
black holes, have been studied for these radiations [15-25]. The
wave equations of spin-1/2 and spin-0 particles, with or without
charge, on three-dimensional black holes such as the charged BTZ
spacetimes \cite{Banados:1992wn, Banados:1992gq, Martinez:1999qi,
Carlip:1991zk}, have also been analyzed \cite{Ejaz:2013fla}, and
the emission and absorption probability of these particles,
incorporating WKB approximation and spacetime symmetries were
investigated. The BTZ spacetimes can also appear from the near
horizon geometry of higher dimensional black holes in appropriate
limits, and can be embedded in supergravity and superstring
theory, for example \cite{Balasubramanian:2007bs,
Fareghbal:2008ar}. In a related approach, it has also been shown
that the propagation of waves described by the Dirac equation of
spin-1/2 fermions in Kerr-Newman spacetime decays in time
\cite{Yau, Finster:2000jz}, and the probability that the spin-1/2
particle escapes to infinity was also computed \cite{Yau}.
Fermions with various spins in a curved spacetime have also been
studied \cite{Yale:2008kx}.

In this paper we study the spin-3/2 particles in accelerating black hole
spacetimes. We first solve a set of Rarita-Schwinger equations of neutral
spin-3/2 particles in the background of accelerating black holes with charge
and rotation. An overall phase factor of the wavefunction for the spin-3/2
particles with given energy and angular momentum can be evaluated. The
probability of emission and absorption of spin-3/2 particles across the
event horizons are computed. After that, we consider charged spin-3/2
particles on the background of charged accelerating black holes, and a
similar analysis are performed. We then analyze the wavefunctions of the
spin-3/2 particles near the acceleration horizon, where the emission and
absorption probability of the spin-3/2 particles are also computed.

This paper is organized as follows. After an brief introduction of the
accelerating charged and rotating black holes in the next Section \ref%
{sec_accelerating bh}, we discuss spin-3/2 particles on the accelerating
black holes in Section \ref{sec_spin 3/2 on accelerating bh}. After this,
charged spin-3/2 particles on charged accelerating black holes are discussed
in Section \ref{sec_charged_particle}. In Section \ref{sec_acceleration
horizon} we discuss the acceleration horizon and the emissions of spin-3/2
particles through the acceleration horizon and the temperature associated
with it. Finally in Section \ref{sec_discussion} we make brief conclusions
of our paper with some discussion.


\section{Accelarating and charged black holes}

\label{sec_accelerating bh}

We consider a family of spacetimes which include an acceleration
parameter \cite{Griffiths:2006tk, Griffiths:2005mi}. It contains
the well known spacetimes like Schwarzchild,
Reissner-Nordstr\"{o}m, Kerr, Kerr-Newman black holes, and many
others as its special cases. It also includes accelerating and
rotating black holes with zero cosmological constant. Nonzero
cosmological constant can also be introduced
\cite{Griffiths:2006tk, Griffiths:2005mi}. The metric for these
black holes in spherical polar coordinates $(t,r,\theta ,\phi )$\
can be written as \cite{Griffiths:2006tk, Griffiths:2005mi}
\begin{eqnarray}
ds^{2} &=&\frac{1}{\Omega ^{2}}\{-(\frac{Q}{\rho ^{2}}-\frac{a^{2}P\sin
^{2}\theta }{\rho ^{2}})dt^{2}+\frac{\rho ^{2}}{Q}dr^{2}+\frac{\rho ^{2}}{P}%
d\theta ^{2}  \nonumber \\
&&+(\frac{P(r^{2}+a^{2})^{2}\sin ^{2}\theta }{\rho ^{2}}-\frac{Qa^{2}\sin
^{4}\theta }{\rho ^{2}})d\phi ^{2}\}-\frac{2a\sin ^{2}\theta
(P(r^{2}+a^{2})-Q)dtd\phi }{\rho ^{2}\Omega ^{2}},  \nonumber \\
&&  \label{bh_01}
\end{eqnarray}
in which
\begin{eqnarray}
\Omega &=&1-\alpha r\cos \theta ,\text{ }  \label{Omega_01} \\
\rho ^{2} &=&r^{2}+a^{2}\cos ^{2}\theta ,  \label{rho_01} \\
P &=&1-2\alpha M\cos \theta +\left[ \alpha ^{2}\left(
e^{2}+g^{2}+a^{2}\right) \right] \cos ^{2}\theta ,  \label{P_01} \\
\text{ }Q &=&\left[ \left( a^{2}+e^{2}+g^{2}\right) -2Mr+r^{2}\right] \left(
1-\alpha ^{2}r^{2}\right) .  \label{Q_01}
\end{eqnarray}%
Here $M$ is the mass of the black hole, $e$ and $g$ are its electric and
magnetic charges, $a$ is the rotation, and $\alpha $ is the acceleration of
the black hole. In Eq. (\ref{bh_01}), rearranging the terms, we get%
\begin{equation}
ds^{2}=-f(r,\theta )dt^{2}+\frac{dr^{2}}{v(r,\theta )}+\Xi (r,\theta
)d\theta ^{2}+K(r,\theta )d\phi ^{2}-2\Phi (r,\theta )dtd\phi ,
\end{equation}%
where $f(r,\theta ),\text{\ }{v(r,\theta )},\text{\ }\Xi (r,\theta ),\text{\
}K(r,\theta )\text{ and\ }\Phi (r,\theta )$ are defined below
\begin{eqnarray}
f(r,\theta ) &=&\frac{1}{\Omega ^{2}}(\frac{Q-a^{2}P\sin ^{2}\theta }{\rho
^{2}}),  \label{f_01} \\
v(r,\theta ) &=&\frac{Q\Omega ^{2}}{\rho ^{2}},  \label{v_01} \\
\Xi (r,\theta ) &=&\frac{\rho ^{2}}{P\Omega ^{2}},  \label{XI_01} \\
K(r,\theta ) &=&\left( \frac{\sin ^{2}\theta \left[
P(r^{2}+a^{2})^{2}-Qa^{2}\sin ^{2}\theta \right] }{\rho ^{2}\Omega ^{2}}%
\right) ,  \label{K_01} \\
\Phi (r,\theta ) &=&\left( \frac{a\sin ^{2}\theta \left[ P(r^{2}+a^{2})-Q%
\right] }{\rho ^{2}\Omega ^{2}}\right) .  \label{PHI_01}
\end{eqnarray}%
The vector potential for these black holes is%
\begin{equation}
A=\frac{-er\left[ dt-a\sin ^{2}\theta d\phi \right] -g\cos \theta \left[
adt-\left( r^{2}+a^{2}\right) d\phi \right] }{r^{2}+a^{2}\cos ^{2}\theta }.
\label{vector_01}
\end{equation}%
These solutions can be obtained in Einstein gravity with Maxwell field.

The horizons are obtained by taking $Q=0$, which gives their locations at
\begin{equation}
r=\frac{1}{\alpha },\text{ \ and \ }r_{\pm }=M\pm \sqrt{%
M^{2}-e^{2}-g^{2}-a^{2}}.  \label{r_horizon_01}
\end{equation}%
Here $r_{\pm }$ represent the outer and inner horizons similar to those of
the Kerr-Newman black holes. We are only considering the case that the sign
inside the radical is always positive. The other horizon at $r=\frac{1}{%
\alpha}$ is an acceleration horizon. In our notations we assume $\alpha
\geqslant 0$.

Now we define the function which will be needed later,
\begin{equation}
F(r,\theta )=f(r,\theta )+\frac{\Phi ^{2}(r,\theta )}{K(r,\theta )}.
\label{F_01}
\end{equation}%
Using the values of $f(r,\theta ),\text{\ }K(r,\theta )\text{ and\ }\Phi
(r,\theta )$ from Eqs. (\ref{f_01}), (\ref{K_01}) and (\ref{PHI_01}) and
after simplification we get
\begin{equation}
F(r,\theta )=\frac{PQ\rho ^{2}}{\left[ P(r^{2}+a^{2})^{2}-Qa^{2}\sin
^{2}\theta \right] \Omega ^{2}}.  \label{3.20}
\end{equation}%
The angular velocity, for the metric (\ref{bh_01}), takes the form
\begin{equation}
\Omega _{H}=\left. \frac{-g_{t\phi }}{g_{\phi \phi }}\right\vert
_{r=r_{+}}=\left. \frac{\Phi (r,\theta )}{K(r,\theta )}\right\vert
_{r=r_{+}}.
\end{equation}
Using the values of $K(r_{+},\theta )\text{ and\ }\Phi (r_{+},\theta )\text{%
\ from Eqs. }$(\ref{K_01}) and (\ref{PHI_01}) we get
\begin{equation}
\Omega _{H}=\frac{a}{r_{+}^{2}+a^{2}}
\end{equation}%
which is evaluated at the outer horizon.


\section{Spin-3/2 particles on accelerating black holes}

\label{sec_spin 3/2 on accelerating bh}

In this section we consider the accelerating charged and rotating black
holes. We consider spin-3/2 fields and their physical properties on the
background of these black hole spacetimes. The Rarita-Schwinger equation of
the spin 3/2 fermion field is of the form \cite{Rarita Schwinger}%
\begin{equation}
i\gamma ^{\nu }(D_{\nu })\Psi _{\mu }-\frac{m}{\hbar }\Psi _{\mu }=0,
\label{R-S_01}
\end{equation}
where $\Psi _{\mu }$ =$\Psi _{\mu \sigma }$ is a vector-valued spinor, with
a vector index and a spinor index, and $m$ is the mass of the field. The $%
D_{\nu }$ is the covariant derivative. The first equation
(\ref{R-S_01}) is the Dirac equation applied to every vector index
of $\Psi _{\mu \sigma }$, while there is a second equation
\cite{Rarita Schwinger}
\begin{equation}
\gamma ^{\mu }\Psi _{\mu }=0,  \label{R-S_02}
\end{equation}%
which is a set of additional constraints. There is a supplementary condition
$D^{\mu }\Psi _{\mu }=0$, which can be derived from these above two
equations \cite{Rarita Schwinger}. These constraints ensure that $\Psi _{\mu
\sigma }$ represents spin-3/2 fermion fields. The field $\Psi _{\mu \sigma }$
describes the spin-3/2 particle and its anti-particle. The curved space $%
\gamma $-matrices are defined as%
\begin{eqnarray}
\gamma ^{t} &=&\sqrt{\frac{(P(r^{2}+a^{2})^{2}-Qa^{2}\sin ^{2}\theta
)(\Omega ^{2})}{PQ\rho ^{2}}}\gamma ^{0},\text{\ \ \ \ \ \ }\gamma ^{r}=%
\sqrt{\frac{Q\Omega ^{2}}{\rho ^{2}}}\gamma ^{3},\text{\ \ \ \ \ \ }\gamma
^{\theta }=\sqrt{\frac{P\Omega ^{2}}{\rho ^{2}}}\gamma ^{1},  \nonumber \\
\gamma ^{\phi } &=&\frac{\rho \Omega \gamma ^{2}}{\sin \theta \sqrt{%
P(r^{2}+a^{2})^{2})-Qa^{2}\sin ^{2}\theta }}+\frac{a(P(r^{2}+a^{2})-Q)\gamma
^{0}}{\sqrt{F(r,\theta )}(P(r^{2}+a^{2})^{2})-Qa^{2}\sin ^{2}\theta )},
\label{3.27}
\end{eqnarray}%
and we choose the basis for the tangent space $\gamma $-matrices to be,%
\begin{equation}
\gamma ^{0}=\left(
\begin{array}{cc}
0 & I_{2} \\
I_{2} & 0%
\end{array}%
\right) ,\text{\ \ \ \ \ \ \ }\gamma ^{i}=\left(
\begin{array}{cc}
0 & \sigma ^{i} \\
-\sigma ^{i} & 0%
\end{array}%
\right) ,  \nonumber
\end{equation}%
with $\{\gamma ^{\alpha },\gamma ^{\beta }\}=-2\eta ^{\alpha \beta }$.

We use the following ansatz for the wave function
\begin{equation}
\Psi _{\mu }(t,r,\theta ,\phi )=\left(
\begin{array}{c}
u_{\mu }^{(1)} \\
u_{\mu }^{(2)} \\
u_{\mu }^{(3)} \\
u_{\mu }^{(4)}
\end{array}
\right) exp\left[ \frac{i}{\hbar }I(t,r,\theta ,\phi )\right] ,
\label{Psi_mu_01}
\end{equation}
in which $u_{\mu }^{(1)},u_{\mu }^{(2)},u_{\mu }^{(3)},u_{\mu }^{(4)}$ each
are functions of the spacetime coordinates.

The first Rarita-Schwinger equation (\ref{R-S_01}) will give an equation
which can be solved for the action $I$ independently of the vector
components of the wave function. The second Rarita-Schwinger equation (\ref%
{R-S_02}) will give four constraints for these vector components of the wave
function independently of the action. Hence, as the action is all we require
to find the horizon temperature, equation (\ref{R-S_02}) will never have any
effect on the action $I.$ This implies that fermions of every spin will emit
at the same temperature.

The covariant derivatives are
\begin{equation}
D_{\mu }=\partial _{\mu }+\frac{1}{8}\omega _{\mu \alpha \beta }[\gamma
^{\alpha },\gamma ^{\beta }],
\end{equation}
where $[\gamma ^{\alpha },\gamma ^{\beta }]$ satisfies the commutative
relations
\begin{equation}
\lbrack \gamma ^{\alpha },\gamma ^{\beta }]=-[\gamma ^{\beta
},\gamma ^{\alpha }],\text{\ \ \ \ \ }if\text{\ \ \ }\alpha \neq
\beta ;\text{\ \ \ \ }[\gamma ^{\alpha },\gamma ^{\beta
}]=0,\text{\ \ \ }if\text{\ \ \ }\alpha =\beta.
\label{commutator_01}
\end{equation}

By using Eq. (\ref{commutator_01}) the Rarita-Schwinger equation takes the
form
\begin{equation}
(i\gamma ^{t}\partial _{t}+i\gamma ^{r}\partial _{r}+i\gamma ^{\theta
}\partial _{\theta }+i\gamma ^{\phi }\partial _{\phi })\Psi _{\mu }-\frac{m}{%
\hbar }\Psi _{\mu }=0.  \label{RS_03}
\end{equation}
Now, we substitute the above ansatz (\ref{Psi_mu_01}) of the wave function
into Eq. (\ref{RS_03}) and compute it term by term. We divide by the
exponential term and neglect the terms with higher orders in $\hbar $. We
obtain the following four equations, for $\mu =0$,$...$,$3$,
\begin{eqnarray}
0 &=&(-\frac{1}{\sqrt{F(r,\theta )}}\partial _{t}I-\sqrt{v(r,\theta )}%
\partial _{r}I-\frac{\Phi (r,\theta )}{K(r,\theta )\sqrt{F(r,\theta )}}%
\partial _{\phi }I)u_{\mu }^{(3)}  \nonumber \\
&&+(-\frac{1}{\sqrt{\Xi (r,\theta )}}\partial _{\theta }I+\frac{i}{\sqrt{%
K(r,\theta )}}\partial _{\phi }I)u_{\mu }^{(4)}-u_{\mu }^{(1)}m, \\
0 &=&(-\frac{1}{\sqrt{F(r,\theta )}}\partial _{t}I+\sqrt{v(r,\theta )}%
\partial _{r}I-\frac{\Phi (r,\theta )}{K(r,\theta )\sqrt{F(r,\theta )}}%
\partial _{\phi }I)u_{\mu }^{(4)}  \nonumber \\
&&-(\frac{1}{\sqrt{\Xi (r,\theta )}}\partial _{\theta }I+\frac{i}{\sqrt{%
K(r,\theta )}}\partial _{\phi }I)u_{\mu }^{(3)}-u_{\mu }^{(2)}m, \\
0 &=&(\frac{1}{\sqrt{F(r,\theta )}}\partial _{t}I-\sqrt{v(r,\theta )}%
\partial _{r}I+\frac{\Phi (r,\theta )}{K(r,\theta )\sqrt{F(r,\theta )}}%
\partial _{\phi }I)u_{\mu }^{(1)}  \nonumber \\
&&+(-\frac{1}{\sqrt{\Xi (r,\theta )}}\partial _{\theta }I+\frac{i}{\sqrt{%
K(r,\theta )}}\partial _{\phi }I)u_{\mu }^{(2)}+u_{\mu }^{(3)}m, \\
0 &=&(\frac{1}{\sqrt{F(r,\theta )}}\partial _{t}I+\sqrt{v(r,\theta )}%
\partial _{r}I+\frac{\Phi (r,\theta )}{K(r,\theta )\sqrt{F(r,\theta )}}%
\partial _{\phi }I)u_{\mu }^{(2)}  \nonumber \\
&&-(\frac{1}{\sqrt{\Xi (r,\theta )}}\partial _{\theta }I+\frac{i}{\sqrt{%
K(r,\theta )}}\partial _{\phi }I)u_{\mu }^{(1)}+u_{\mu }^{(4)}m.
\end{eqnarray}%
The second set of Rarita-Schwinger equation (\ref{R-S_02}) yields%
\begin{eqnarray}
0 &=&\frac{u_{t}^{(3)}}{\sqrt{F}}+\sqrt{v}u_{r}^{(3)}+\frac{u_{\theta }^{(4)}%
}{\sqrt{\Xi }}+\frac{\Phi u_{\phi }^{(3)}}{K\sqrt{F}}-\frac{iu_{\phi }^{(4)}%
}{\sqrt{K}},  \label{RS_02_01} \\
0 &=&\frac{u_{t}^{(4)}}{\sqrt{F}}-\sqrt{v}u_{r}^{(4)}+\frac{u_{\theta }^{(3)}%
}{\sqrt{\Xi }}+\frac{\Phi u_{\phi }^{(4)}}{K\sqrt{F}}+\frac{iu_{\phi }^{(3)}%
}{\sqrt{K}}, \\
0 &=&\frac{-u_{t}^{(1)}}{\sqrt{F}}+\sqrt{v}u_{r}^{(1)}+\frac{u_{\theta
}^{(2)}}{\sqrt{\Xi }}-\frac{\Phi u_{\phi }^{(1)}}{K\sqrt{F}}-\frac{iu_{\phi
}^{(2)}}{\sqrt{K}}, \\
0 &=&\frac{-u_{t}^{(2)}}{\sqrt{F}}-\sqrt{v}u_{r}^{(2)}+\frac{u_{\theta
}^{(1)}}{\sqrt{\Xi }}-\frac{\Phi u_{\phi }^{(2)}}{K\sqrt{F}}+\frac{iu_{\phi
}^{(1)}}{\sqrt{K}}.  \label{RS_02_04}
\end{eqnarray}%
These will give us additional relation between the various vector components
of the wave function, but will not influence the common phase factor given
by $I~$in Eq. (\ref{Psi_mu_01}), so these are not important here as the
solution for the action will be independent of these relations.

Taking into account the symmetries of the spacetime at hand, we employ the
ansatz for the action as
\begin{equation}
I=-Et+J\phi +W(r,\theta ).  \label{action_01}
\end{equation}%
Here $E$ and $J$ denote the energy and angular momentum of the radiated
particle. We denote $\frac{\partial W}{\partial r}~$as $W_{r}$ for
simplicity.~Inserting this into the above four equations, we get%
\begin{eqnarray}
0 &=&(\frac{E}{\sqrt{F(r,\theta )}}-\sqrt{v(r,\theta )}W_{r}-\frac{J\Phi
(r,\theta )}{K(r,\theta )\sqrt{F(r,\theta )}})u_{\mu }^{(3)}  \nonumber \\
&&+(-\frac{1}{\sqrt{\Xi (r,\theta )}}W_{\theta }+\frac{iJ}{\sqrt{K(r,\theta )%
}})u_{\mu }^{(4)}-u_{\mu }^{(1)}m,  \label{RS_EJ_01} \\
0 &=&(\frac{E}{\sqrt{F(r,\theta )}}+\sqrt{v(r,\theta )}W_{r}-\frac{J\Phi
(r,\theta )}{K(r,\theta )\sqrt{F(r,\theta )}})u_{\mu }^{(4)}  \nonumber \\
&&-(\frac{1}{\sqrt{\Xi (r,\theta )}}W_{\theta }+\frac{iJ}{\sqrt{K(r,\theta )}%
})u_{\mu }^{(3)}-u_{\mu }^{(2)}m, \\
0 &=&(-\frac{E}{\sqrt{F(r,\theta )}}-\sqrt{v(r,\theta )}W_{r}+\frac{J\Phi
(r,\theta )}{K(r,\theta )\sqrt{F(r,\theta )}})u_{\mu }^{(1)}  \nonumber \\
&&+(-\frac{1}{\sqrt{\Xi (r,\theta )}}W_{\theta }+\frac{iJ}{\sqrt{K(r,\theta )%
}})u_{\mu }^{(2)}+u_{\mu }^{(3)}m, \\
0 &=&(-\frac{E}{\sqrt{F(r,\theta )}}+\sqrt{v(r,\theta )}W_{r}+\frac{J\Phi
(r,\theta )}{K(r,\theta )\sqrt{F(r,\theta )}})u_{\mu }^{(2)}  \nonumber \\
&&-(\frac{1}{\sqrt{\Xi (r,\theta )}}W_{\theta }+\frac{iJ}{\sqrt{K(r,\theta )}%
})u_{\mu }^{(1)}+u_{\mu }^{(4)}m.  \label{RS_EJ_04}
\end{eqnarray}%
Expanding Eq. (\ref{v_01}) in Taylor's series and neglecting the higher
powers we get
\begin{equation}
v(r,\theta )=v(r_{+},\theta )+(r-r_{+})v_{r}(r_{+},\theta ),
\end{equation}%
where we denote $\frac{\partial v}{\partial r}~$as $v_{r}$, and similarly
for other functions. Note that at the horizon $v(r_{+},\theta )=0,$ and its
derivative at the horizon is $v_{r}(r_{+},\theta )$, so it becomes%
\begin{equation}
v(r,\theta )=(r-r_{+})(\frac{2(r_{+}-M)(1-\alpha ^{2}r_{+}^{2})\Omega
^{2}\left( r_{+},\theta \right) }{(r_{+}^{2}+a^{2}\cos ^{2}\theta )}).
\label{v_near horizon}
\end{equation}%
Similarly, we expand $F(r,\theta )=F(r_{+},\theta
)+(r-r_{+})F_{r}(r_{+},\theta )$,$~$and noting that at the horizon $%
F(r_{+},\theta )=0$, Eq. (\ref{F_01}) becomes
\begin{equation}
F(r,\theta )=(r-r_{+})(\frac{2(r_{+}^{2}+a^{2}\cos ^{2}\theta
)(r_{+}-M)(1-\alpha ^{2}r_{+}^{2})}{(r_{+}^{2}+a^{2})^{2}\Omega ^{2}\left(
r_{+},\theta \right) }).  \label{F_near horizon}
\end{equation}%
Expanding near the black hole horizon, from Eqs. (\ref{RS_02_01}) to (\ref%
{RS_02_04}), we obtain four equations which have no effects on the solution
of the action. Now expanding Eqs. (\ref{RS_EJ_01}) to (\ref{RS_EJ_04}) near
the black hole horizon and using Eqs. (\ref{v_near horizon}) and (\ref%
{F_near horizon}), we get the first Rarita-Schwinger equation in the
simplified form,%
\begin{eqnarray}
\left[
\begin{array}{cccc}
-m & 0 & \varepsilon -W_{r}\sqrt{(r-r_{+})v_{r}} & s-\frac{W_{\theta }}{%
\sqrt{\Xi }} \\
0 & -m & -s-\frac{W_{\theta }}{\sqrt{\Xi }} & \varepsilon +W_{r}\sqrt{%
(r-r_{+})v_{r}} \\
\varepsilon +W_{r}\sqrt{(r-r_{+})v_{r}} & -s+\frac{W_{\theta }}{\sqrt{\Xi }}
& -m & 0 \\
s+\frac{W_{\theta }}{\sqrt{\Xi }} & \varepsilon -W_{r}\sqrt{(r-r_{+})v_{r}}
& 0 & -m%
\end{array}%
\right] \left[
\begin{array}{c}
u_{\mu }^{(1)} \\
u_{\mu }^{(2)} \\
u_{\mu }^{(3)} \\
u_{\mu }^{(4)}%
\end{array}%
\right] &=&0,  \nonumber \\
&&
\end{eqnarray}%
where we define%
\begin{eqnarray}
\varepsilon &=&\frac{E}{\sqrt{(r-r_{+})F_{r}(r_{+},\theta )}}-\frac{J~\Phi }{%
K\sqrt{(r-r_{+})F_{r}(r_{+},\theta )}}, \\
s &=&i\frac{J}{\sqrt{K}}.
\end{eqnarray}%
We make a Left-Upper Decomposition of the matrix and this gives,%
\begin{equation}
(\varepsilon +W_{r}\sqrt{(r-r_{+})v_{r}})(\varepsilon -W_{r}\sqrt{%
(r-r_{+})v_{r}})-(s+\frac{W_{\theta }}{\sqrt{\Xi }})(-s+\frac{W_{\theta }}{%
\sqrt{\Xi }})-m^{2}=0.  \label{a}
\end{equation}%
It can be seen that we can take $\theta =\theta _{0}$, where $\theta _{0}$
is constant of motion, and then $W_{\theta }$ can be solved to be a
constant. Then expanding Eq. (\ref{a}) near the horizon $r=r_{+}$ and
solving for $W_{r}$, we see that only $\varepsilon $ contributes, as the
absolute value of $W_{r}$ increases quickly to infinity near the horizon
while the other terms do not contribute in the solution. Near the horizon we
see that $W(r,\theta )$ can be written as$~W(r)+\xi (\theta )$,~where $\xi $
is complex and has no $r$ dependence. We see that the $W_{r}$ has$~$two
solutions, $\partial _{r}W_{+}$ and $\partial _{r}W_{-}$, corresponding to
outgoing and incoming modes respectively,
\begin{equation}
\partial _{r}W_{\pm }=\pm \frac{(E-\Omega _{H}J)}{\sqrt{%
(r-r_{+})F_{r}(r_{+},\theta )}\sqrt{(r-r_{+})v_{r}(r_{+},\theta )}}.
\end{equation}%
Substituting the values of $F_{r}(r_{+},\theta )$ and $v_{r}(r_{+},\theta )$%
, the above equations becomes
\begin{equation}
\partial _{r}W_{\pm }=\pm \frac{(E-\Omega _{H}J)(r_{+}^{2}+a^{2})}{%
2(r-r_{+})(r_{+}-M)(1-\alpha ^{2}r_{+}^{2})}.
\end{equation}%
The $\partial _{r}W_{\pm }$ has a pole at the horizon $r_{+}$. For finding
the value of $W$ we integrate the above result
\begin{equation}
W_{\pm }=\pm \int \frac{(E-\Omega _{H}J)(r_{+}^{2}+a^{2})dr}{%
2(r-r_{+})(r_{+}-M)(1-\alpha ^{2}r_{+}^{2})}.
\end{equation}%
Integrating the above integrand around the pole $r=r_{+}$, this gives
\begin{equation}
W_{\pm }=\pm \pi i\frac{(E-\Omega _{H}J)(r_{+}^{2}+a^{2})}{%
2(r_{+}-M)(1-\alpha ^{2}r_{+}^{2})}
\end{equation}%
and we obtain
\begin{equation}
\func{Im}W_{\pm }=\pm \frac{\pi }{2}\frac{(E-\Omega _{H}J)(r_{+}^{2}+a^{2})}{%
(r_{+}-M)(1-\alpha ^{2}r_{+}^{2})}.
\end{equation}%
So the probabilities of the emission and absorption of the spin-3/2
particles are
\begin{eqnarray}
P_{emission} &\propto &exp[-2\func{Im}I_{+}]=exp[-2(\func{Im}W_{+}+\func{Im}%
\xi )], \\
P_{absorption} &\propto &exp[-2\func{Im}I_{-}]=exp[-2(\func{Im}W_{-}+\func{Im%
}\xi )].
\end{eqnarray}%
Since $\func{Im}W_{+}=-\func{Im}W_{-}$,
\begin{equation}
\Gamma =\frac{P_{emission}}{P_{absorption}}=\frac{exp[-2(\func{Im}W_{+})]}{%
exp[-2(\func{Im}W_{-})]}=exp[-4\func{Im}W_{+}].
\end{equation}%
The resulting tunneling probability is
\begin{equation}
\Gamma =exp[-2\pi \frac{(E-\Omega _{H}J)(r_{+}^{2}+a^{2})}{%
(r_{+}-M)(1-\alpha ^{2}r_{+}^{2})}]=exp[-\beta (E-\Omega _{H}J)].
\label{probability_01}
\end{equation}%
Comparing this with $\beta =1/T_{H}$\ we find that the horizon temperature
is given by%
\begin{equation}
T_{H}=\frac{(r_{+}-M)(1-\alpha ^{2}r_{+}^{2})}{2\pi (r_{+}^{2}+a^{2})},
\label{temperature_01}
\end{equation}%
where $r_{+}$ is given by Eq. (\ref{r_horizon_01}). If we set rotation and
acceleration equal to zero, these expressions will give the expressions for
the Reissner-Nordstr\"{o}m black hole. If we compare it with the black hole
without acceleration, we note that the effect of acceleration is that it
decreases the temperature.



\section{The charged case}

\label{sec_charged_particle}

In this section we consider charged spin-3/2 particles on the accelerating
and rotating charged black hole, with the outer horizon $r_{+}=M+\sqrt{%
M^{2}-e^{2}-g^{2}-a^{2}}$. The Rarita-Schwinger equation of the spin 3/2
fermion $\Psi _{\mu }$ with charge $q$ is given by
\begin{equation}
i\gamma ^{\nu }\left( D_{\nu }-\frac{iq}{\hbar }A_{\nu }\right) \Psi _{\mu }-%
\frac{m}{\hbar }\Psi _{\mu }=0,
\end{equation}%
where $q$ and $m$ is the charge and mass of the particle and $A_{\mu }$ is
the vector potential.

Using an ansatz similar to the one in Section \ref{sec_spin 3/2 on
accelerating bh}, the above wave equation takes the form%
\begin{equation}
i\gamma ^{t}\partial _{t}\Psi _{\mu }+i\gamma ^{r}\partial _{r}\Psi _{\mu
}+i\gamma ^{\theta }\partial _{\theta }\Psi _{\mu }+i\gamma ^{\phi }\partial
_{\phi }\Psi _{\mu }+\gamma ^{t}\frac{q}{\hbar }A_{t}\Psi _{\mu }+\gamma
^{\phi }\frac{q}{\hbar }A_{\phi }\Psi _{\mu }-\frac{m}{\hbar }\Psi _{\mu }=0.
\end{equation}%
Substituting the functions as in Section \ref{sec_accelerating bh}
and using the vector potential $A_{\mu }$ given by Eq.
(\ref{vector_01}), after simplification, we obtain four equations,
\begin{eqnarray}
0 &=&-[\frac{1}{\sqrt{F\left( r,\theta \right) }}\partial _{t}I+\sqrt{%
v\left( r,\theta \right) }\partial _{r}I+\frac{\Phi \left( r,\theta \right)
}{K\left( r,\theta \right) \sqrt{F\left( r,\theta \right) }}\partial _{\phi
}I-\frac{1}{\sqrt{F\left( r,\theta \right) }}qA_{t}  \nonumber \\
&&-\frac{\Phi \left( r,\theta \right) }{K\left( r,\theta \right) \sqrt{%
F\left( r,\theta \right) }}qA_{\phi }]u_{\mu }^{(3)}+[-\frac{1}{\sqrt{\Xi
\left( r,\theta \right) }}\partial _{\theta }I+\frac{i}{\sqrt{K\left(
r,\theta \right) }}\partial _{\phi }I-\frac{i}{\sqrt{K\left( r,\theta
\right) }}qA_{\phi }]u_{\mu }^{(4)}-u_{\mu }^{(1)}m,  \nonumber \\
&&  \label{RS_cha_01} \\
0 &=&-[\frac{1}{\sqrt{F\left( r,\theta \right) }}\partial _{t}I-\sqrt{%
v\left( r,\theta \right) }\partial _{r}I+\frac{\Phi \left( r,\theta \right)
}{K\left( r,\theta \right) \sqrt{F\left( r,\theta \right) }}\partial _{\phi
}I-\frac{1}{\sqrt{F\left( r,\theta \right) }}qA_{t}  \nonumber \\
&&-\frac{\Phi \left( r,\theta \right) }{K\left( r,\theta \right) \sqrt{%
F\left( r,\theta \right) }}qA_{\phi }]u_{\mu }^{(4)}-[\frac{1}{\sqrt{\Xi
\left( r,\theta \right) }}\partial _{\theta }I+\frac{i}{\sqrt{K\left(
r,\theta \right) }}\partial _{\phi }I-\frac{i}{\sqrt{K\left( r,\theta
\right) }}qA_{\phi }]u_{\mu }^{(3)}-u_{\mu }^{(2)}m,  \nonumber \\
&& \\
0 &=&[\frac{1}{\sqrt{F\left( r,\theta \right) }}\partial _{t}I-\sqrt{v\left(
r,\theta \right) }\partial _{r}I+\frac{\Phi \left( r,\theta \right) }{%
K\left( r,\theta \right) \sqrt{F\left( r,\theta \right) }}\partial _{\phi }I-%
\frac{1}{\sqrt{F\left( r,\theta \right) }}qA_{t}  \nonumber \\
&&-\frac{\Phi \left( r,\theta \right) }{K\left( r,\theta \right) \sqrt{%
F\left( r,\theta \right) }}qA_{\phi }]u_{\mu }^{(1)}+[-\frac{1}{\sqrt{\Xi
\left( r,\theta \right) }}\partial _{\theta }I+\frac{i}{\sqrt{K\left(
r,\theta \right) }}\partial _{\phi }I-\frac{i}{\sqrt{K\left( r,\theta
\right) }}qA_{\phi }]u_{\mu }^{(2)}+u_{\mu }^{(3)}m,  \nonumber \\
&& \\
0 &=&[\frac{1}{\sqrt{F\left( r,\theta \right) }}\partial _{t}I+\sqrt{v\left(
r,\theta \right) }\partial _{r}I+\frac{\Phi \left( r,\theta \right) }{%
K\left( r,\theta \right) \sqrt{F\left( r,\theta \right) }}\partial _{\phi }I-%
\frac{1}{\sqrt{F\left( r,\theta \right) }}qA_{t}  \nonumber \\
&&-\frac{\Phi \left( r,\theta \right) }{K\left( r,\theta \right) \sqrt{%
F\left( r,\theta \right) }}qA_{\phi }]u_{\mu }^{(2)}-[\frac{1}{\sqrt{\Xi
\left( r,\theta \right) }}\partial _{\theta }I+\frac{i}{\sqrt{K\left(
r,\theta \right) }}\partial _{\phi }I-\frac{i}{\sqrt{K\left( r,\theta
\right) }}qA_{\phi }]u_{\mu }^{(1)}+u_{\mu }^{(4)}m.  \nonumber \\
&&  \label{RS_cha_04}
\end{eqnarray}%
From Eq. (\ref{RS_02_01}) to Eq. (\ref{RS_02_04}), we obtained
four equations which have no effects on the solution of the
action.

In order to solve these equations (\ref{RS_cha_01}) to
(\ref{RS_cha_04}), we again employ an action of the form given in
Eq. (\ref{Psi_mu_01}) and substitute it in the above equations.
Near the horizon\ at $r=r_{+}$, the functions take the following
form,
\begin{eqnarray}
F\left( r,\theta \right) &=&\left( r-r_{+}\right) \partial _{r}F\left(
r_{+},\theta \right)  \nonumber \\
&=&\left( r-r_{+}\right) \frac{\left( r_{+}^{2}+a^{2}\cos ^{2}\theta \right)
\left( 2r_{+}-2M\right) \left( 1-\alpha ^{2}r_{+}^{2}\right) }{\left(
1-\alpha r_{+}\cos \theta \right) ^{2}\left( r_{+}^{2}+a^{2}\right) ^{2}}, \\
v\left( r,\theta \right) &=&\left( r-r_{+}\right) \partial _{r}v\left(
r_{+},\theta \right)  \nonumber \\
&=&\left( r-r_{+}\right) \frac{\left( 1-\alpha r_{+}\cos \theta \right)
^{2}\left( 2r_{+}-2M\right) \left( 1-\alpha ^{2}r_{+}^{2}\right) }{\left(
r_{+}^{2}+a^{2}\cos ^{2}\theta \right) }, \\
\Omega _{H} &=&\frac{\Phi \left( r_{+},\theta \right) }{K\left( r_{+},\theta
\right) }=\frac{a}{\left( r_{+}^{2}+a^{2}\right) }.
\end{eqnarray}%
Using these values in Eq. (\ref{RS_cha_01}) to Eq. (\ref{RS_cha_04}) and
expanding near the horizon we obtain%
\begin{eqnarray}
0 &=&(\varepsilon -W_{r}\sqrt{(r-r_{+})v_{r}(r_{+},\theta )})u_{\mu
}^{(3)}+(s-\frac{W_{\theta }}{\sqrt{\Xi (r_{+},\theta )}})u_{\mu
}^{(4)}-u_{\mu }^{(1)}m,  \label{RS_cha_simplified_01} \\
0 &=&(\varepsilon +W_{r}\sqrt{(r-r_{+})v_{r}(r_{+},\theta )})u_{\mu
}^{(4)}+(-s-\frac{W_{\theta }}{\sqrt{\Xi (r_{+},\theta )}})u_{\mu
}^{(3)}-u_{\mu }^{(2)}m, \\
0 &=&(\varepsilon +W_{r}\sqrt{(r-r_{+})v_{r}(r_{+},\theta )})u_{\mu
}^{(1)}-(-s-\frac{W_{\theta }}{\sqrt{\Xi (r_{+},\theta )}})u_{\mu
}^{(2)}-u_{\mu }^{(3)}m, \\
0 &=&(\varepsilon -W_{r}\sqrt{(r-r_{+})v_{r}(r_{+},\theta )})u_{\mu
}^{(2)}-(s-\frac{W_{\theta }}{\sqrt{\Xi (r_{+},\theta )}})u_{\mu
}^{(1)}-u_{\mu }^{(4)}m,  \label{RS_cha_simplified_04}
\end{eqnarray}
where we define
\begin{eqnarray}
\varepsilon &=&\frac{\left( E-\Omega _{H}J-\frac{qer_{+}}{\left(
r_{+}^{2}+a^{2}\right) }\right) }{\sqrt{\left( r-r_{+}\right) \partial
_{r}F\left( r_{+},\theta \right) }}, \\
s &=&i(\frac{J}{\sqrt{K}}+\frac{qer}{\rho ^{2}\sqrt{K}}a\sin ^{2}\theta ).
\end{eqnarray}%
Near the horizon, we also have the functions as following,
\begin{eqnarray}
\Xi (r_{+},\theta ) &=&\frac{\rho ^{2}\left( r_{+},\theta \right) }{P\Omega
^{2}\left( r_{+},\theta \right) }, \\
K\left( r_{+},\theta \right) &=&\left( \frac{\sin ^{2}\theta P\left(
r_{+}^{2}+a^{2}\right) ^{2}}{\rho ^{2}\left( r_{+},\theta \right) \Omega
^{2}\left( r_{+},\theta \right) }\right) , \\
\Phi \left( r_{+},\theta \right) &=&\left( \frac{a\sin ^{2}\theta P\left(
r_{+}^{2}+a^{2}\right) }{\rho ^{2}\left( r_{+},\theta \right) \Omega
^{2}\left( r_{+},\theta \right) }\right) .
\end{eqnarray}

Solving the above equations (\ref{RS_cha_simplified_01}) to (\ref%
{RS_cha_simplified_04}), as we discussed in Section \ref{sec_spin 3/2 on
accelerating bh}, finally we get%
\begin{equation}
\text{\ \ \ \ }W_{\pm }=\pm \frac{\pi i\left( E-\Omega _{H}J-\frac{qer_{+}}{%
\left( r_{+}^{2}+a^{2}\right) }\right) \left( r_{+}^{2}+a^{2}\right) }{%
\left( 2r_{+}-2M\right) \left( 1-\alpha ^{2}r_{+}^{2}\right) },
\end{equation}%
and
\begin{equation}
\func{Im}W_{\pm }=\pm \frac{\pi }{2}\frac{\left( E-\Omega _{H}J-\frac{qer_{+}%
}{\left( r_{+}^{2}+a^{2}\right) }\right) (r_{+}^{2}+a^{2})}{%
(r_{+}-M)(1-\alpha ^{2}r_{+}^{2})}.
\end{equation}%
So the tunneling probability becomes%
\begin{equation}
\Gamma =\frac{P_{emission}}{P_{absorption}}=exp[-4\func{Im}W_{+}].
\end{equation}%
Using the value of $\func{Im}W_{+}$ in the above equation we obtain%
\begin{eqnarray}
\Gamma &=&exp[-\frac{2\pi (r_{+}^{2}+a^{2})}{(r_{+}-M)(1-\alpha
^{2}r_{+}^{2})}\left( E-\Omega _{H}J-\frac{qer_{+}}{\left(
r_{+}^{2}+a^{2}\right) }\right) ]  \label{tunnel_charge_02} \\
&=&exp[-\beta (E-\Omega _{H}J-V_{H}q)],
\end{eqnarray}%
which is the Boltzmann factor of the emitted charged particles, including
the chemical potential conjugate to the charge $q$. Comparing this
expression with $\beta =1/T_{H}$, this gives the horizon temperature as%
\begin{equation}
T_{H}=\frac{(r_{+}-M)(1-\alpha ^{2}r_{+}^{2})}{2\pi (r_{+}^{2}+a^{2})},
\label{temperature_02}
\end{equation}%
and $r_{+}$ in this case is given by Eq. (\ref{r_horizon_01}). If
we put acceleration and rotation equal to zero in formulae
(\ref{tunnel_charge_02}) and (\ref{temperature_02}), they reduce
to the tunneling probability and the temperature of the
Reissner-Nordstr\"{o}m black hole. If we set the charges of the
background to be zero, they also recover the expressions in
Section \ref{sec_spin 3/2 on accelerating bh}.


\section{The acceleration horizon and the charged rotating $C$-metric}

\label{sec_acceleration horizon}

The charged $C$-metric can give a description of a Reissner-Nordstr\"{o}m
black hole with constant proper acceleration and with mass $M$ and charge $e$%
,$~$for example \cite{Griffiths:2006tk, Corn Utt}. In the extended space
with the coordinate system adapted to boost-rotation symmetry, the complete
space can describe a pair of charged black holes with uniform accelerations
in opposite directions, for example \cite{Griffiths:2006tk,
Griffiths:2005mi, Bicak, Pravda:2000zm, Corn Utt}. It is an example of
boost-rotation symmetric spacetime, with an axial Killing vector and a boost
Killing vector. The spin-3/2 particles can be emitted through the horizons
of these backgrounds.

In the metric of these backgrounds (\ref{bh_01}), the range of $\theta $ is $%
\theta \in \lbrack 0,\pi ]$. \vspace{1pt}The function $\Omega =1-\alpha
r\cos \theta $, has no zero for $\theta \in \lbrack \frac{\pi }{2},\pi ]$.
Due to the $\frac{1}{\Omega ^{2}}$ factor of the metric, for $\theta \in
\lbrack 0,\frac{\pi }{2}]$, the conformal infinity is located at the zero of
the function $\Omega $, which is $r=\frac{1}{\alpha \cos \theta }~$\cite%
{Griffiths:2006tk, Griffiths:2005mi}.

The horizons of these spacetimes are obtained by setting $Q=0$ in Eq. (\ref%
{Q_01}), which have three positive roots. The two roots%
\begin{equation}
r_{\pm }=M\pm \sqrt{M^{2}-e^{2}-g^{2}-a^{2}},  \label{outer_inner_horizon}
\end{equation}%
represent the outer and inner horizons. We are only considering the case
that the sign inside the radical is always positive. The other root is
\begin{equation}
r_{a}=\frac{1}{\alpha },  \label{acceleration_horizon}
\end{equation}%
which is an acceleration horizon related to a boost symmetry, and the
subscript $a$ in $r_{a}$ denotes acceleration.

Now we consider a different solution of the function $W$ for the spin-3/2
particles defined in Eq. (\ref{action_01}) in Section \ref{sec_spin 3/2 on
accelerating bh}. In Sections \ref{sec_spin 3/2 on accelerating bh} and \ref%
{sec_charged_particle}, we have the solution where the pole of $W_{r}$ is at
$r=r_{+}$. Now we look for a different solution where the pole of $W_{r}$ is
at $r=r_{a}$. We then get the set of equations in the form of Eq. (\ref%
{RS_cha_01}) to Eq. (\ref{RS_cha_04}). Near the acceleration horizon, both $%
F(r,\theta )$ and $v(r,\theta )$ have a zero at $r_{a}$, so we can expand
them as $F(r,\theta )=F(r_{a},\theta )+(r-r_{a})F_{r}(r_{a},\theta )$ and $%
v(r,\theta )=v(r_{a},\theta )+(r-r_{a})v_{r}(r_{a},\theta )$. \vspace{1pt}We
then get four equations in the similar form as Eq. (\ref%
{RS_cha_simplified_01}) to Eq. (\ref{RS_cha_simplified_04}) in Section \ref%
{sec_charged_particle}, except that in this section we expand the functions
at a different zero of the function $Q$. The $W_{r}$ has$~$two solutions, $%
\partial _{r}W_{+}$ and $\partial _{r}W_{-}$, corresponding to outgoing and
incoming modes,
\begin{eqnarray}
\partial _{r}W_{\pm } &=&\pm \frac{(E-\Omega _{a}J-\frac{qe\alpha }{%
1+a^{2}\alpha ^{2}})}{\sqrt{(r-r_{a})F_{r}}\sqrt{(r-r_{a})v_{r}}} \\
&&  \nonumber \\
&=&\pm \frac{(1+a^{2}\alpha ^{2})(E-\Omega _{a}J-\frac{qe\alpha }{%
1+a^{2}\alpha ^{2}})}{2\alpha (r-r_{a})(1-2M\alpha +\left(
e^{2}+g^{2}+a^{2}\right) \alpha ^{2})},
\end{eqnarray}%
where $\Omega _{a}=\frac{a}{r_{a}^{2}+a^{2}}$.~By integrating the above
integrand around the pole at $r=r_{a}$, we get%
\begin{equation}
W_{\pm }=\pm i\frac{\pi (1+a^{2}\alpha ^{2})(E-\Omega _{a}J-\frac{qe\alpha }{%
1+a^{2}\alpha ^{2}})}{2\alpha (1-2M\alpha +\left( e^{2}+g^{2}+a^{2}\right)
\alpha ^{2})},
\end{equation}%
and
\begin{equation}
\Gamma =exp[-4\func{Im}W_{+}]=exp[-\beta (E-\Omega _{a}J-\frac{qe\alpha }{%
1+a^{2}\alpha ^{2}})].
\end{equation}%
Comparing this expression with $\beta =1/T$, we get the temperature of the
emitted spin-3/2 particles at the acceleration horizon,%
\begin{equation}
T_{a}=\frac{\alpha (1-2M\alpha +\left( e^{2}+g^{2}+a^{2}\right) \alpha ^{2})%
}{2\pi (1+a^{2}\alpha ^{2})},  \label{acceleration_temperature}
\end{equation}%
where the subscript $a$ in $T_{a}~$denotes acceleration.

Now we look at the surface gravity at the horizons. There is a coordinate
singularity at the black hole event horizon $r_{+}~$and at the acceleration
horizon $r_{a}$. These can be removed by using Painlev\'{e}-type coordinate
transformation. We perform the coordinate transformation,%
\begin{equation}
dt\rightarrow dt-\sqrt{\frac{1-v}{Fv}}dr,
\end{equation}%
so that the Eq. (\ref{bh_01}) becomes,%
\begin{equation}
ds^{2}=-F(r,\theta )dt^{2}+dr^{2}+2F\sqrt{\frac{1-v}{Fv}}drdt+\Xi (r,\theta
)d\theta ^{2}+K(r,\theta )d{\bar{\phi}}^{2},  \label{metric_transformed}
\end{equation}%
where $d{\bar{\phi}}=d\phi -\frac{\Phi (r,\theta )}{K(r,\theta )}dt$. In
this form the surface gravity becomes%
\begin{equation}
\kappa \mid _{r=r_{\ast }}=\left. \frac{1}{2}\sqrt{\frac{1-v}{Fv}}v\frac{dF}{%
dr}\right\vert _{r=r_{\ast }},  \label{surface_gravity}
\end{equation}%
where $r_{\ast }$ denotes the locations of the horizons, the black hole
horizon $r_{+}$ and the acceleration horizon $r_{a}$.~

Using this formula (\ref{surface_gravity}) for the $C$-metric Eq. (\ref%
{metric_transformed}), we obtain result at the black hole event horizon,%
\begin{equation}
\kappa \mid _{r=r_{+}}=\frac{(r_{+}-M)(1-\alpha ^{2}r_{+}^{2})}{%
r_{+}^{2}+a^{2}},
\end{equation}%
where $r_{+}$ is the outer horizon in Eq. (\ref{outer_inner_horizon}). The
temperature at the black hole horizon, $T=\kappa /2\pi $, of the $C$-metric
agrees with the computation in Eq. (\ref{temperature_02}). \vspace{1pt}%
\vspace{1pt}By using Eq. (\ref{surface_gravity}), the surface gravity at the
acceleration horizon is,
\begin{equation}
\kappa \mid _{r=r_{a}}=\frac{\alpha (1-2M\alpha +\left(
e^{2}+g^{2}+a^{2}\right) \alpha ^{2})}{1+a^{2}\alpha ^{2}}.
\end{equation}%
The surface gravity at the acceleration horizon of the $C$-metric thus
agrees with the computation of Eq. (\ref{acceleration_temperature}), that is
$\kappa \mid _{r=r_{a}}=2\pi T_{a}$.


\section{Discussion}

\label{sec_discussion}

We have considered the Rarita-Schwinger equations for spin-3/2
particles on the backgrounds of accelerating black holes with
general electric charge, magnetic charge, rotation, and
acceleration parameter. We first consider the neutral spin-3/2
particles. The overall phase factor in the vector components of
the wavefunctions for the spin-3/2 particle are analyzed and its
radial derivative has poles on the locations of the event
horizons. By the integration along complex path we evaluated the
emission and absorption probabilities of the spin-3/2 particles
across the horizons, as well as the temperatures of the horizons.

We also analyzed the charged spin-3/2 particles, on the background
of these accelerating black holes, with parameters of electric and
magnetic charges and rotation. The spin-3/2 wave equations are
analyzed similarly for this case. The Boltzmann factor of the
emitted outgoing particles are derived, including the chemical
potential conjugate to the charge of the particle.

Because the spacetime here is a black hole with acceleration, there are two
types of event horizons. One is the black hole event horizon at the outer
horizon radius, and another is the acceleration horizon due to the
acceleration of the source. By the integration along complex path we
obtained also the emission and absorption probabilities of the spin-3/2
particles across the acceleration horizon.

The properties of the wavefunctions of the spin-3/2 particles near the
acceleration horizon are also analyzed in detail. The surface gravity at the
acceleration horizon calculated after performing a coordinate
transformation, matches with the temperature calculated by the
Rarita-Schwinger equations.


\acknowledgments

This work was supported in part by NSF grant DMS-1159412, NSF grant
PHY-0937443, NSF grant DMS-0804454, and in part by the Fundamental Laws
Initiative of the Center for the Fundamental Laws of Nature, Harvard
University.


\end{document}